**Title: A fast radio burst localised to a massive galaxy**

Authors: V. Ravi[1,2,3,*], M. Catha[4], L. DAddario[1], S. G. Djorgovski[1], G. Hallinan[1], R. Hobbs[4], J. Kocz[1], S. R. Kulkarni[1], J. Shi[1], H. K. Vedantham[1,5], S. Weinreb[1] & D. P. Woody[4]

[1]Cahill Center for Astronomy and Astrophysics, MC 249-17, California Institute of Technology, Pasadena CA 91125, USA.

[2]Center for Astrophysics | Harvard & Smithsonian, 60 Garden Street, Cambridge MA 02138, USA.

[3]E-mail: vikram@caltech.edu.

[4]Owens Valley Radio Observatory, California Institute of Technology, Big Pine CA 93513, USA.

[5]ASTRON, Netherlands Institute for Radio Astronomy, Oude Hoogeveensedijk 4, 7991PD, Dwingeloo, The Netherlands.

**Intense, millisecond-duration bursts of radio waves have been detected from beyond the Milky Way[1]. Their extragalactic origins are evidenced by their large dispersion measures, which are greater than expected for propagation through the Milky Way interstellar medium alone, and imply contributions from the intergalactic medium and potentially host galaxies[2]. Although several theories exist for the sources of these fast radio bursts, their intensities, durations and temporal structures suggest coherent emission from highly magnetised plasma[3,4].**

**Two sources have been observed to repeat[5,6], and one repeater (FRB 121102) has been localised to the largest star-forming region of a dwarf galaxy at a cosmological redshift of 0.19 [Refs. 7, 8]. However, the host galaxies and distances of the so far non-repeating fast radio bursts are yet to be identified. Unlike repeating sources, these events must be observed with an interferometer with sufficient spatial resolution for arcsecond localisation at the time of discovery. Here we report the localisation of a fast radio burst (FRB 190523) to a few-arcsecond region containing a single massive galaxy at a redshift of 0.66. This galaxy is in stark contrast to the host of FRB 121102, being a thousand times more massive, with a greater than hundred times lower specific star-formation rate. The properties of this galaxy highlight the possibility of a channel for FRB production associated with older stellar populations.**

We detected the fast radio burst (FRB) 190523 on the modified Julian date (MJD) 58626.254118233(2) with the Deep Synoptic Array ten-antenna prototype (DSA-10, see Methods). Throughout this Letter, we quote the standard errors (68% confidence) in the least significant figures in parentheses. The DSA-10 consists of 4.5-m radio dishes separated by 6.75 m to 1300 m, located at the Owens Valley Radio Observatory. The DSA-10 is designed to detect FRBs in the phase-incoherent combination of signals from each dish, and to then process the same signals interferometrically (coherent combination) to localise FRBs to few-arcsecond accuracy. FRB 190523 was detected at a dispersion measure (DM) of 760.8(6) pc $cm^{-3}$, and localised to the coordinates (J2000): right ascension (RA) 13:48:15.6(2), declination (DEC) +72:28:11(2). A time-frequency data set was formed at this position through the coherent addition of interferometric visibility data from DSA-10 (see Methods). These data, displayed in Figure 1, consist of total-intensity spectra recorded in 1250 frequency channels between 1334.69 MHz and 1487.28 MHz over 131.072 μs intervals, with the data in each channel incoherently corrected with at least 8.192 μs accuracy for the dispersive delay. The burst signal to noise ratio exceeds 10 in multiple time samples. The observed properties of FRB 190523 are summarised in Table 1. We derive a fluence of approximately 280 Jy ms given the sensitivity of DSA-10 at the burst location within the field of view. No repeat bursts at this position were detected during approximately 78 hr of observations obtained over 54 days surrounding the detection (see Methods).

The 99% confidence containment region of FRB 190523 (Figure 2) includes just one galaxy in archival data from the Panoramic Survey Telescope and Rapid Response System (Pan-STARRS) 3π Steradian Survey[9]. This galaxy, PSO J207.0643+72.4708 (hereafter PSO J207+72), was detected with an *r*-band magnitude of 22.1(1) in the stacked Pan-STARRS data. We obtained images of the containment region of the burst on MJD 58635 with the Keck I telescope of the W. M. Keck Observatory, using the Low Resolution Imaging Spectrometer (KeckI/LRIS, see Methods)[10]. No objects besides PSO J207+72 were detected within the FRB 190523 99% confidence containment region to limiting magnitudes of 25.8 in the *g* filter and 26.1 in the *R* filter. The containment region lies within an apparent grouping of galaxies (Figure 2), with the galaxy nearest to the containment region (S2 in Figure 2) detected by Pan-STARRS with an *r*-band magnitude of 22.1(1). We also obtained a low-resolution optical spectrum of PSO J207+72 using KeckI/LRIS on MJD 58635 (see Methods). The spectrum (Figure 3) indicates stellar absorption features at a redshift of 0.660(2). A single emission line, corresponding to the [OII] 3727 Angstrom doublet, is tentatively detected with a flux of 4.7(7) x $10^{-17}$ erg $s^{-1}$ $cm^{-2}$.

We model the Pan-STARRS optical photometry and the KeckI/LRIS optical spectroscopy of PSO J207+72 using the Prospector software[11,12]. We used this software to fit a 'delay-tau' stellar population and star-formation history model to the data. In this model, the star-formation history is proportional to $te^{-t/\tau}$, where $t$ is the time since the formation epoch of the galaxy, and $\tau$ is the characteristic decay time of the star-formation history. We derive a metallicity fraction of 0.3(2) of the solar metallicity, a stellar mass of $10^{11.07(6)}$ $M_\odot$ (where $M_\odot$ is a solar mass), and an ongoing star-formation rate of approximately 1.3 $M_\odot$ $yr^{-1}$. The star-formation rate, although poorly constrained given the limited wavelength coverage of the data, is consistent with the flux of the possible [OII] 3727 Angstrom emission line, which implies a star-formation rate of up to 1.3(2) $M_\odot$ $yr^{-1}$ [Ref. 13]. As this emission line could also be associated with weak activity of the central black hole[14], we adopt a star-formation rate of 1.3(2) $M_\odot$ $yr^{-1}$ as an approximate upper limit for PSO J207+72.

As the only object detected within the containment region of FRB 190523, PSO J207+72 is the likely host galaxy of the burst. Additional evidence is furnished by the agreement between the burst DM and the predicted DM for the redshift of PSO J207+72. Accounting for 37 pc $cm^{-3}$ from the Milky Way disk[15], and between 50 and 80 pc $cm^{-3}$ from the Milky Way ionised halo[16], the extragalactic DM of FRB 190523 is between 644 and 674 pc $cm^{-3}$. Given this extragalactic DM, and parameterising the containment region by the 3 x 8 arcsec 95% confidence ellipse[17], the probability of finding any galaxy (even one not detectable in our data) by chance within the containment region is <10% [Ref. 17]. Further, the redshift of PSO J207+72 is not larger than expected given the DM of FRB 190523. The DM contributed by the intergalactic medium (IGM) to the redshift of PSO J207+72 is $660(f_{\rm IGM}/0.7)$ pc cm$^{-3}$, where $f_{\rm IGM}$ is the fraction of the luminous matter (termed baryons) of the Universe in the ionised IGM[18]. Observations suggest that 60% of cosmic baryons are in the IGM ($f_{\rm IGM} \approx 0.6(1)$), 10% of baryons are locked in galaxies, and that the remaining 30% of baryons are apportioned between the circum-galactic medium in galaxy halos and the IGM[19]. We adopt $f_{\rm IGM} = 0.7$ as a fiducial value, noting that an rms scatter of approximately 200 pc cm$^{-3}$ in the IGM DM to redshifts around 0.66 is expected due to cosmic variance and intervening galaxy halos[20]. Finally, our *R*-band KeckI/LRIS image excludes the possibility that the FRB 190523 containment region includes a dwarf galaxy like the host of the repeating FRB 121102 at a redshift below approximately 0.45 (2.5 Gpc luminosity distance)[8]. The <10% probability of chance coincidence of the burst containment region with a galaxy, even one as small as the FRB 121102 host, implies that there is a <10% probability that the FRB 190523 containment region includes a galaxy like the FRB 121102 host. This further suggests that PSO J207+72 is the unique host galaxy of FRB 190523.

The properties of FRB 190523 are typical of FRBs observed at frequencies around 1.4 GHz [Ref. 21]. At the distance of PSO J207+72, FRB 190523 has a spectral energy of $5.6 \times 10^{33}$ erg Hz$^{-1}$, which is consistent with the largest previously estimated burst energies[22]. The patchy spectrum of FRB 190523 (Figure 1) is also similar to the spectra of bright FRBs detected by the Australian Square Kilometre Array Pathfinder[22]. We note that our DSA-10 observations cannot exclude the possibility of repeated bursts from FRB 190523 below our detection threshold, which FRB 190523 itself only exceeded by 15%.

The temporal profile of FRB 190523 indicates a broadening timescale due to multi-path propagation through inhomogeneous plasma of 1.4(2) ms at 1 GHz [Ref. 21]. This broadening timescale is three orders of magnitude higher than expected for the sightline of FRB 190523 through the Milky Way interstellar medium[15]. The broadening timescale is also higher than expected during propagation through the DM column potentially contributed by PSO J207+72 ($\lesssim$150 pc cm$^{-3}$)[23]. Our results therefore support the possibility of some FRBs being temporally broadened during propagation between their host galaxies and the Milky Way, such as in the circum-galactic medium of intervening galaxies[24].

The properties of PSO J207+72 are in tension with FRB progenitor models developed based on the host galaxy of the repeating FRB 121102 [Ref. 25]. In particular, the host of FRB 121102 is similar to the dwarf star-forming host galaxies of superluminous supernovae and long gamma-ray bursts, which are the terminal explosions of the most massive stars. However, the stellar mass of PSO J207+72 is higher and its star-formation rate per unit mass is lower than the known host galaxies of superluminous supernovae and long gamma-ray bursts at redshifts below 1 [Ref. 25]. In addition, leading models for the FRB emission mechanism favour neutron star progenitors with magnetar magnetic field strengths ($\gtrsim$10$^{14}$ G) [Refs. 3, 4, 26]. If this is the case, our results suggest that magnetars that were formed in the terminal explosions of the most massive stars are not the only such objects capable of emitting FRBs. Indeed, magnetar formation channels exist that do not require young stellar populations, such as the accretion induced collapse of white dwarfs to neutron stars in mass-transfer binaries[27,28] and the merger of two neutron stars[29].

The likely low contribution to the DM of FRB 190523 from PSO J207+72 provides evidence in support of FRB progenitor models (magnetar or not) that do not require actively star-forming environments. The low global star-formation rate of PSO J207+72, and the spatially offset location of much of the containment region of FRB 190523 relative to the galaxy (Figure 2), lead us to consider the possibility that the progenitor of FRB 190523 was drawn from an old stellar population. The similarity between the stellar populations of PSO J207+72 and the Milky Way implies that galaxies like the Milky Way can harbour FRB progenitors.

**Table 1: Properties of FRB 190523 and its host galaxy.**

| Property | Measurement |
|---|---|
| Topocentric arrival time at 1530 MHz (MJD) | 58626.254118233(2) |
| Fluence (Jy ms) | ≳280 |
| Dispersion measure (pc cm$^{-3}$) | 760.8(6) |
| Dispersion measure index | -2.003(8) |
| Milky Way disk (halo) dispersion measure (pc cm$^{-3}$) | 37 (50 to 80) |
| Extragalactic dispersion measure (pc cm$^{-3}$) | 644 to 674 |
| Band averaged full-width half-maximum of the burst (ms) | 0.42(5) |
| Scattering timescale at 1 GHz (ms) | 1.4(2) |
| Right ascension (J2000) | 13:48:15.6(2) |
| Declination (J2000) | +72:28:11(2) |
| Host galaxy redshift | 0.660(2) |
| Host galaxy luminosity distance (Gpc) | 4.08(1) |
| Burst energy (erg Hz$^{-1}$) | 5.6 x $10^{33}$ |
| Host galaxy stellar mass ($M_\odot$) | $10^{11.07(6)}$ |
| Host galaxy star-formation rate ($M_\odot$ yr$^{-1}$) | <1.3 |

**Supplementary Information** is linked to the online version of the paper at www.nature.com/nature.

**Acknowledgements.** We thank the staff of the Owens Valley Radio Observatory, including J. Lamb, K. Hudson, A. Rizo and M. Virgin, for their assistance with the construction of the DSA-10. We thank A. Readhead for supporting the initiation of the DSA-10 project. We also thank A. Soliman for assistance with the development of the DSA-10 receivers. A portion of this research was performed at the Jet Propulsion Laboratory, California Institute of Technology, under a President and Directors Fund grant and under a contract with the National Aeronautics and Space Administration. This research was additionally supported by the National Science Foundation under grant AST-1836018. VR acknowledges support as a Millikan Postdoctoral Scholar in Astronomy at the California Institute of Technology, and from a Clay Postdoctoral Fellowship of the Smithsonian Astrophysical Observatory. SGD acknowledges a partial support from the NSF grant AST-1815034 and the NASA grant 16-ADAP16-0232. Some of the data presented herein were obtained at the W. M. Keck Observatory, which is operated as a scientific partnership among the California Institute of Technology, the University of California and the National Aeronautics and Space Administration. The Observatory was made possible by the generous financial support of the W. M. Keck Foundation. This research made use of Astropy, a community-developed core Python package for Astronomy. The Pan-STARRS1 Surveys (PS1) and the PS1 public science archive have been made possible through contributions by the Institute for Astronomy, the University of Hawaii, the Pan-STARRS Project Office, the Max-Planck Society and its participating institutes, the Max Planck Institute for Astronomy, Heidelberg and the Max Planck Institute for Extraterrestrial Physics, Garching, The Johns Hopkins University, Durham University, the University of Edinburgh, the Queen's University Belfast, the Harvard-Smithsonian Center for Astrophysics, the Las Cumbres Observatory Global Telescope Network Incorporated, the National Central University of Taiwan, the Space Telescope Science Institute, the National Aeronautics and Space Administration under Grant No. NNX08AR22G issued through the Planetary Science Division of the NASA Science Mission Directorate, the National Science Foundation Grant No. AST-1238877, the University of Maryland, Eotvos Lorand University (ELTE), the Los Alamos National Laboratory, and the Gordon and Betty Moore Foundation.

**Author contributions.** G.H., V.R. and H.K.V. conceived of and developed the DSA-10 concept and observing strategy. V.R., J.K., and S.R.K. led the construction and initial deployment of DSA-10. D.W., S.W, L.D., J.K., V.R., H.K.V, M.C. and R.H. designed and built the DSA-10 subsystems. V.R. and H.K.V. commissioned the DSA-10. V.R. operated DSA-10 and analysed the data. S.G.D. carried out the optical observations. V.R. analysed the optical data, and led the writing of the manuscript with the assistance of all co-authors.

**Author information.** Reprints and permissions information is available at www.nature.com/reprints. The authors declare no competing interests. Correspondence and requests for materials should be addressed to vikram@caltech.edu.

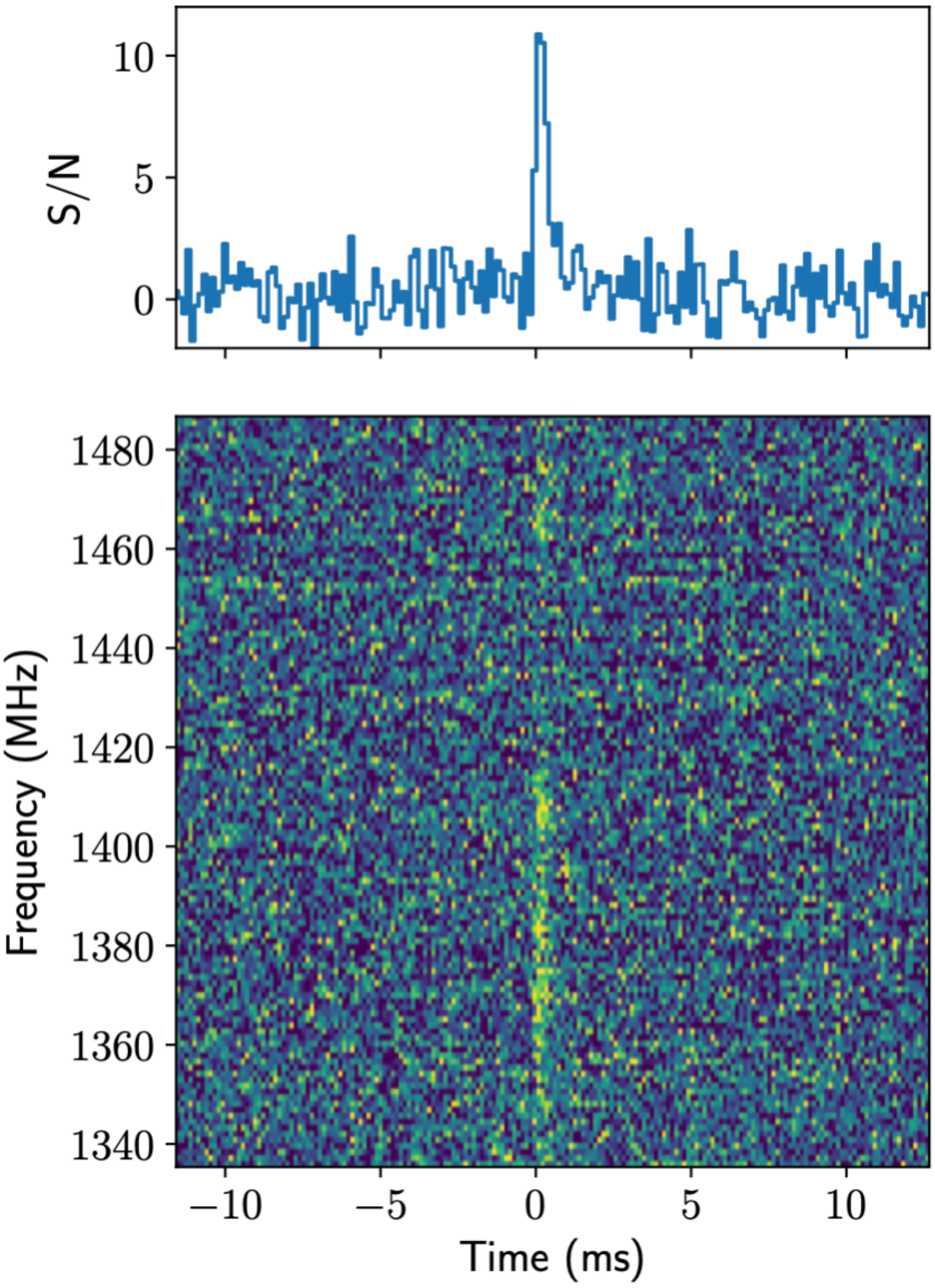

**Figure 1: Time-frequency data on FRB 190523.** The top panel shows the de-dispersed temporal profile of the burst averaged over the DSA-10 frequency band. The data are measures of the received power in 131.072 μs bins, in units of the rms off-burst noise. The bottom panel shows the de-dispersed dynamic spectrum of the burst, again in units of the rms off-burst noise in each 1.22 MHz frequency channel. Although the structure evident in the burst spectrum is likely to be qualitatively accurate, no calibration of the relative flux-density scales in different frequency channels has been applied.

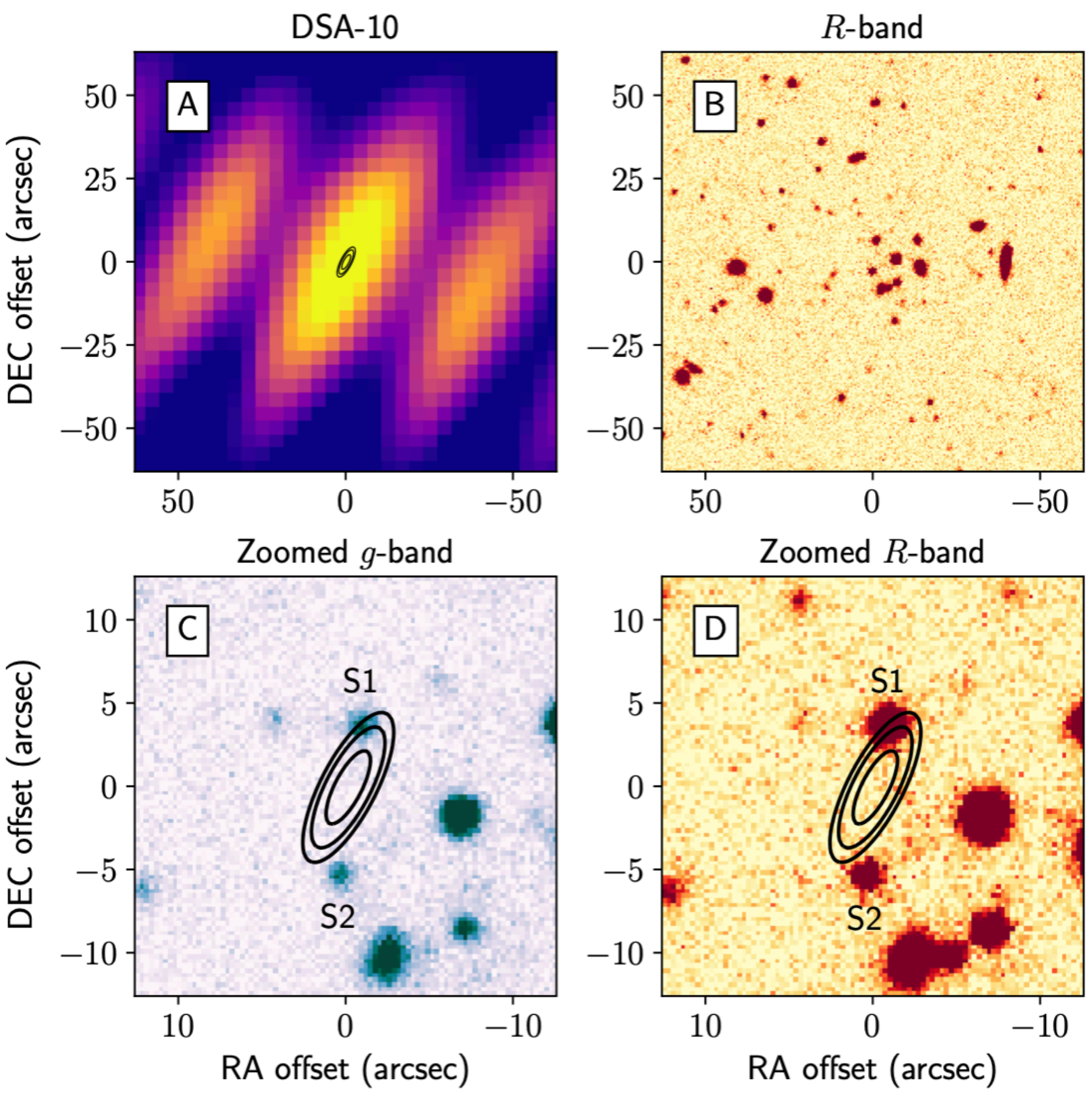

**Figure 2: Images of the sky location of FRB 190523.** All images are centred on co-ordinates (J2000) RA 13:48:15.6, DEC +72:28:11. Panel A shows a dirty snapshot image of the burst obtained with DSA-10 (see Methods). Panel B shows an optical image in the *R*-band filter obtained with KeckI/LRIS. The position of FRB 190523 coincides with an apparent grouping of galaxies. Panels C and D show the zoomed burst localisation region in the *g* and *R* filters of KeckI/LRIS. The position of FRB 190523 is indicated with 68%, 95% and 99% confidence containment ellipses in Panels A, C and D. The only galaxy detected above the 26.1-magnitude *R*-band detection limit within the 99% confidence containment ellipse, indicated by 'S1', is PSO J207+72. A galaxy to the south of the 99% confidence ellipse is labelled `S2'.

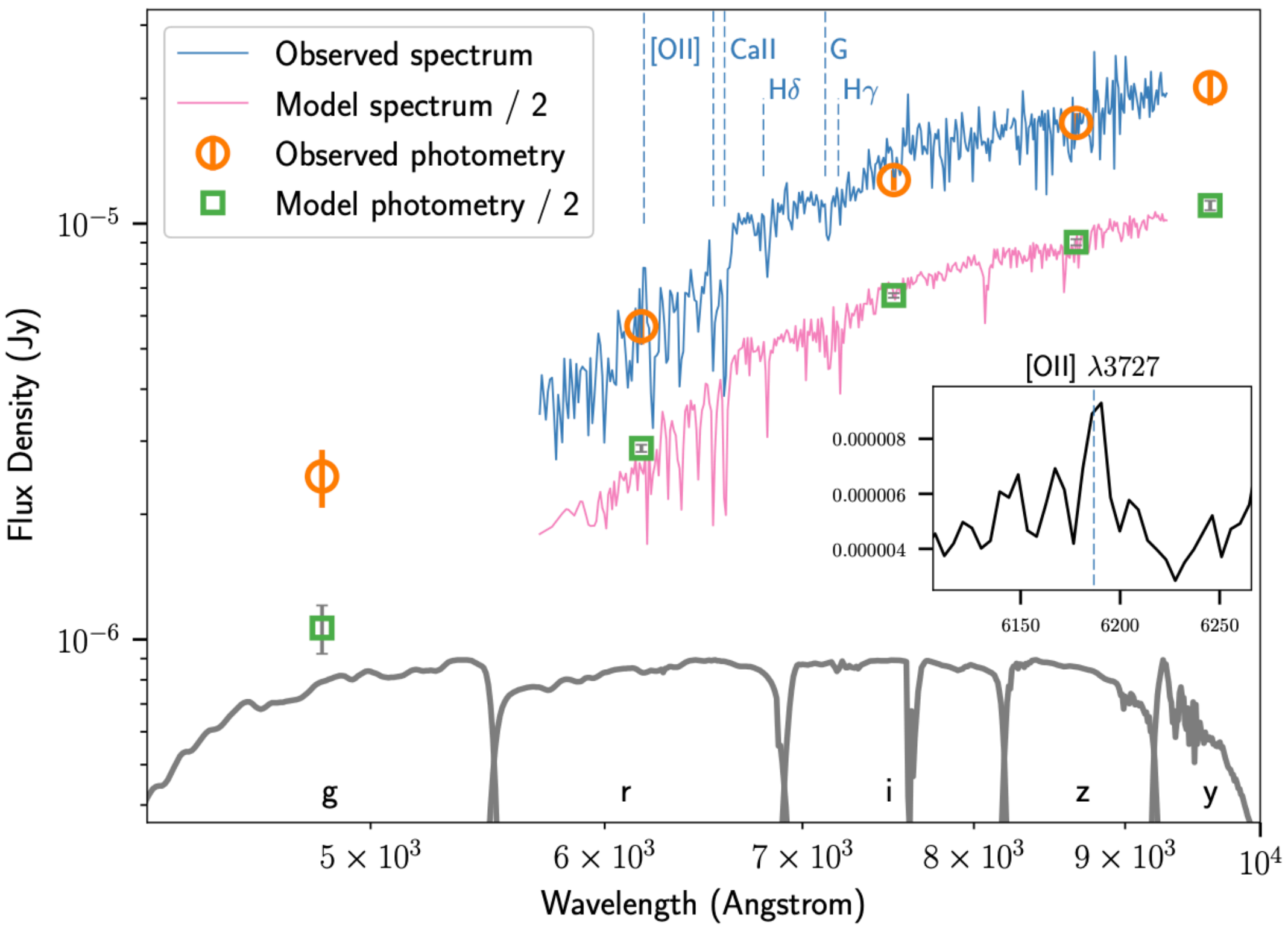

**Figure 3: Modelling of the host galaxy of FRB 190523.** The grey curves illustrate the transmissions of the *g*, *r*, *i*, *z* and *y* Pan-STARRS filters. Error bars indicating one standard deviation are shown for the Pan-STARRS photometry. The maximum aposteriori probability (MAP) results from the Prospector modelling of the host galaxy are scaled downwards by a factor of two. The grey error bars accompanying the MAP photometry points indicate the 5th and 95th percentiles of 500 samples drawn from the posterior parameter distributions. The redshifted positions of some significant absorption lines are indicated by dashed blue traces. The inset shows the observed spectrum around the [OII] 3727 Angstrom feature, binned by a factor of two less than the spectrum in the main panel.

## Methods

**The DSA-10 instrument.** The Deep Synoptic Array 10-element prototype (DSA-10) is an array of ten 4.5-m radio dishes operating in the 1.28 - 1.53 GHz frequency band. The array is deployed at the Owens Valley Radio Observatory (OVRO; located at 37.2314 deg N, 118.2941 deg W) near the town of Bishop, California, USA. A description of the DSA-10 instrument is given in Ref. 30. Here we describe the state of the instrument at the time that FRB 190523 was detected.

The array was in a slightly modified configuration relative to its initial deployment, with four antennas clustered at the northern end of the OVRO Tee-shaped infrastructure. The positions of each antenna, in standard International Terrestrial Reference Frame (ITRF) geocentric coordinates, are given in Extended Data Table 1. Each antenna was equipped with two receivers sensitive to orthogonal linear polarisations. The antenna primary beams have full-width half-maxima of 3.25 deg. Antenna 2 was not operational because it was being used to test new equipment, and one polarisation of Antenna 8 was operating with significantly reduced sensitivity caused by a malfunctioning low-noise amplifier. Antenna 2 was discarded from all calibration and imaging procedures described below. The array operated in a stationary drift-scan mode on the meridian at a declination of +73.6 deg, with an absolute pointing accuracy of better than 0.4 deg. The projected baseline lengths ranged between 5.75 m and 1256.57 m.

The DSA-10 was operated in this configuration between MJD 58568 and MJD 58630, with a total time on sky of 54 days. FRB searching was conducted using the incoherent sum of dynamic spectra from the eight fully functioning antennas, forming a single stream of 2048-channel spectra integrated over 131.072 μs. Prior to summation, the dynamic spectra were excised of narrow-band and impulsive broadband radio-frequency interference (RFI)[30]. These data were searched for FRBs in real time using the Heimdall software[31], with 2477 optimally spaced DM-trials between 30 pc $cm^{-3}$ and 3000 pc $cm^{-3}$. At each trial DM, the data were smoothed with boxcar filters spaced by powers of two between $2^0$ and $2^8$ samples prior to searching. The detection threshold was set at eight standard deviations (8σ). In this Letter, we assume a typical band-averaged system-equivalent flux density of 22 kJy for each DSA-10 receiver, based on interferometric measurements of the system sensitivity using sources with known flux densities[30]. Given eight fully functioning antennas, and 220 MHz of effective bandwidth following RFI excision, this implies an approximate detection threshold of 94 Jy ms at the centre of the primary beam for a millisecond-duration FRB not affected by intra-channel dispersion smearing[32].

Upon detection of any pulse candidate exceeding the detection threshold at any trial DM, 294912 samples of complex voltage data corresponding to each polarisation of each antenna were written to disk. These data consisted of 4-bit real, 4-bit imaginary 2048-channel voltage spectra sampled every 8.192 μs, calculated on and transmitted to five servers by Smart Network ADC Processor (SNAP-1) boards[33] over a 10-Gigabit Ethernet network. The data dumps were extracted from ring buffers such that the candidate pulse arrival times at 1530 MHz were 61035 samples into the dumps.

These voltage data were also used to derive interferometric visibilities between each pair of antennas. The visibilities were measured by integrating the cross-power over 0.402653184 s, and over 625 pairs of channels between exactly 1334.6875 MHz and 1487.275390625 MHz. Approximate, constant path-length delay corrections were digitally applied to each receiver input on the SNAP-1 boards, but no time-dependent fringe tracking corrections were applied online. Visibility data were only recorded when bright unresolved radio sources were transiting through the DSA-10 primary beam. These data were fringe-stopped offline by dividing the data by a model for the visibilities given the known source positions from the NRAO VLA Sky Survey (NVSS) catalogue[34]. Visibility modelling was accomplished using differential antenna positions referenced to the known ITRF location of the centre of the OVRO Tee (which had previously hosted the Caltech OVRO Millimeter Array), using the Common Astronomy Software Applications (CASA, version 5.1.1) package to calculate baseline coordinates. We consider visibility data on three such sources in this Letter: NVSS J120019+730045 (also 3C 268.1, 5.56 Jy; hereafter J1200+7300), NVSS J145907+714019 (7.47 Jy; hereafter J1459+7140), and NVSS J192748+735802 (3.95 Jy; hereafter J1927+7358). Data on these sources were recorded for 3630 s, 1960 s, 3890 s, respectively, centred on their transit times.

**Interferometric calibration and localisation of FRB 190523.** Standard radio-interferometric data processing strategies[35] were used to calibrate the instrumental responses of each DSA-10 antenna and receiver. Here we describe the specific methods used to calibrate the data on FRB 190523, and the steps taken to verify their efficacy.

FRB 190523 was detected on MJD 58626.254118233(2), and a voltage-data dump was successfully triggered. These data were cross-correlated offline using the same routines as applied in the online correlator software[36], and the visibilities were integrated over 131.072 μs. Only data in 1250 channels covering the frequency band (1334.6875 - 1487.275390625 MHz) spanned by the visibility data recorded in real time were retained in the analysis presented here.

At the time FRB 190523 was detected, the DSA-10 pointing centre was at a position (J2000) of RA 14:15:01.98, DEC +73:40 (absolute pointing accuracy of better than 0.4 deg). The calibrator sources J1459+7140 and J1200+7300 transited 29.58 minutes later and 163.76 minutes earlier, respectively. The phases of the per-receiver complex gain corrections for the FRB 190523 data were derived as follows. No attempt at per-receiver gain amplitude calibration was made. This was because all sources under consideration (including FRB 190523) were consistent with unresolved point sources, based on NVSS data[34], that dominated the sky brightness within their fields. All visibility amplitudes were taken to be unity, such that only phase information was preserved.

1. First, receiver based relative delay errors (with Antenna 7 as a reference) were calculated using fringe-stopped data on J1459+7140, restricted to 15 min surrounding transit. J1459+7140 is considered a primary calibrator in the data base of the Very Large Array for baseline lengths consistent with the DSA-10.

2. After applying these delay corrections to the 15 min of J1459+7140 data surrounding transit, the data were averaged in time, and in frequency to 25 channels. The averaged data were used to derive receiver-based phase errors in each channel.

3. The phase solutions from J1459+7140 were averaged with phase solutions derived from 15 min of fringe stopped data on J1200+7300 surrounding transit, with the same delay corrections as above applied first. No significant differences were evident between the phase solutions derived independently from J1459+7140 and J1200+7300.

4. The delay and phase solutions from the above analysis were used to calibrate the visibility data on FRB 190523. The phase centre was set to the array pointing centre at the time of the burst. The data were converted to the Measurement Set (MS) format for further analysis with CASA. Data on the four shortest baselines (after removing baselines with Antenna 2) were excluded because of significant levels of correlated noise. A 7 x 7 deg total-intensity image, without deconvolution of the synthesised beam shape (a 'dirty' image), was then made using four visibility time-samples centred on the burst with the standard imaging task tclean applied for gridding and Fourier inversion. A single point-like source was evident in this image, at a position 2.3 deg from the pointing centre (an hour angle of 26.8 arcmin west, and 1.2 deg south).

5. Given the apparent offset location of the burst from the pointing centre, we then corrected for any direction dependent instrumental-response variations intrinsic to the DSA-10 antennas. This was done by extracting 6 min of fringe-stopped data on J1200+7300 at the same hour angle as the possible position of FRB 190523, applying the previous calibration solutions, and deriving frequency-averaged phase corrections for each receiver (again using Antenna 7 as a reference). We note that no data on J1459+7140 were available at the hour angle of the burst, as visibilities were recorded on this source for a shorter time (1960 s) than for J1200+7300 (3630 s). Significant corrections of up to 25 deg in phase were required, which were identical for the two receivers on each antenna. This formed the final set of calibration solutions for FRB 190523.

We then applied these final calibration solutions to the visibility data on FRB 190523, and referenced the data to a phase centre corresponding to the approximate burst position, with the burst dispersion accounted for in calculating baseline coordinates. The data were then summed over the two polarisations, and converted to the MS format. The CASA task tclean was used to make dirty and deconvolved images of the burst data (see Figure 2, and Extended Data Figure 1 - bottom row). The imaging process was verified using the 6 min of data on J1200+7300 obtained at the same hour angle as FRB 190523 (Extended Data Figure 1 - top row). No sources were detected in images made using visibility data in 128 time samples on either side of FRB 190523, either when averaged together or binned by four samples.

The position of FRB 190523 was estimated by fitting to the calibrated visibilities, using four 131.072 μs time samples centred on the burst, as before. This fit was carried out using the MIRIAD[37] task uvfit (after conversion of the MS to a MIRIAD-format file), and a grid-search code, as a software problem in the CASA task uvmodelfit prevented it from loading our data. The grid-search code was used to evaluate the poste-rior probability of the source position given the likelihood of the visibility data over a uniform 0.25 arcsec grid of positions centred on the burst position in its image. This was then used to calculate the maximum aposteriori probability location quoted in Table 1, and the 68%, 95% and 99% confidence containment ellipses shown in Figure 2 of the main text. We also attempted to estimate the position of FRB 190523 using only data between 1350 MHz and 1420 MHz, where much of the burst spectral energy density appears to be concentrated (Figure 1). This yielded a containment ellipse that was consistent with the result from all the data, but with major and minor axes that were 10% larger.

We verified the efficacy of our localisation procedure using a selection of methods. First, no significant phase closure errors were evident in the calibrated data on J1200+7300 and J1459+7140, either at boresight, or at the hour angle of FRB 190523 in the case of J1200+7300. No baseline-based calibration corrections were required to accurately model the calibrator data. Second, we verified that the calibration solutions derived as above for FRB 190523 were also able to calibrate visibility data on the source J1927+7358, which transited five hours after the burst detection. We did this by extracting 6 min of data on J1927+7358 at the hour angle that the burst was detected, applying the same calibration solutions as applied to the burst, and imaging it and deriving its position as done for the burst (Extended Data Figure 1 - middle row). The position of J1927+7358 was recovered to within 1 arcsec in both dimensions, with the offsets consistent with the position-fit errors. Plots of the calibrated, frequency averaged visibility phases on each baseline of 6 min of data on J1200+7300 and J1927+7358 after rotation to their known positions are shown in Extended Data Fig-ure 2, together with the same results for FRB 190523 rotated to its derived position. The single worst outliers in Extended Data Figure 2 for FRB 190523 and J1927+7358 were both on baselines containing Antenna 1.

We repeated the same calibration procedure as above on the 12 days of data prior to the detection of the burst, and correctly recovered the position of J1927+7358 on each day (Extended Data Figure 3). The rms scatter in the recovered positions of J1927+7358 about the true value was 0.47 arcsec in RA, and 0.69 arcsec in DEC. We therefore have no basis to add a systematic error contribution to the position-fit errors for FRB 190523. We have previously verified that no temporal error existed in the voltage-data dumps by imaging giant pulses from the Crab pulsar (B0531+21) when the DSA-10 was pointed at its declination, by running the same software[30]. We ensured that this remained the case by calibrating and imaging data dumps obtained close to when J1200+7300 was transiting on other days using the above procedures, and verifying that the position of J1200+7300 was correctly recovered.

The astrometric reference for our results was the VLA calibrator catalogue, which was accurate to <0.01 arcsec for J1459+7140 and J1927+7358, and to the NVSS accuracy of approximately 0.5 arcsec [Ref.. 34] for J1200+7300. These errors are small in comparison with the final localisation accuracy of FRB 190523, and hence we do not include them in the localisation error budget of the burst.

**Properties of FRB 190523.** We modelled the temporal profile of FRB 190523 using methods presented in Ref. 21. The data presented in Figure 1 were formed by the co-herent addition of calibrated visibility data on FRB 190523 using its best-fit position. These data were integrated over five evenly spaced frequency bands, and the resulting time series were fit with a series of models. The best-fit model was the convolution of the instrumental response to a delta-function impulse and a one-sided exponential with a timescale varying as $f^{4}$, where $f$ is the observed frequency. This is consistent with temporal broadening caused by multi-path propagation. The extrapolated broadening timescale at 1 GHz is quoted in Table 1. We also quote the uncertainty in the DM index in Table 1; the burst arrival time was found to scale with $f^{2.003(8)}$.

No attempt was made to calibrate the response of the DSA-10 to polarised radiation. The DSA-10 was not designed for polarimetry. First, we have not established our ability to robustly calibrate the per-receiver frequency dependent gain amplitudes using transiting continuum calibrator sources. We also do not record full-polarisation visibility data on these sources, making it impossible to measure the signal leakages between the receivers sensitive to orthogonal linear polarisations. The lack of polarisation information is not likely to affect the burst localisation, because each polarisation was calibrated independently using unpolarised sources. We verified that consistent positions for FRB 190523 were derived for data in each polarisation separately.

**KeckI/LRIS observations and analysis.** KeckI/LRIS observations of the localisation region and candidate host galaxy of FRB 190523 (PSO J207+72) were carried out on the night of MJD 58635 in dark time, under clear photometric conditions with a median seeing-disk full-width half-maximum (*R*-band) of 1.1 arcsec. Light from the telescope was split between the two arms of LRIS by the D560 dichroic. Three images were taken in the *g*- and *R*-band filters at an airmass of 1.66, with exposure times of 30 s, 300 s, and 300 s, and no binning of the detector pixels. The *g*- and *R*-band filters have effective wavelengths of 4731 Angstrom and 6417 Angstrom, respectively, and effective widths of 1082 Angstrom and 1185 Angstrom, respectively. Three spectral exposures were obtained (900-s exposure times, median airmass 1.68) with a 1.5 arc-sec long slit at a position angle of 270 deg, the 600/4000 grism for the blue arm, the 400/8500 grating for the red arm, and the detector binned by two pixels in the spectral direction. The spectral flux-calibration was obtained with observations of the standard star Feige 67 at an airmass of 1.07.

All optical data were reduced using standard procedures for LRIS. Bias subtraction using the overscan levels, flat-fielding using dome-flat exposures, and cosmic-ray rejection was performed with the lpipe software[38]. The imaging data were then astrometrically registered against *Gaia* DR2 stars using the scamp software routines[39], co-added using the swarp software routines[40], and sources extracted using the SExtractor software[41]. Photometric calibration to an accuracy of 0.1 magnitudes was accomplished using Pan-STARRS objects in the field. The weakest detected sources in the *g* and *R* bands were 25.8 and 26.1 magnitudes (AB) respectively, which we adopt as our limiting magnitudes in these bands.

The lpipe software was additionally used to process the spectroscopic data by performing wavelength calibration using internal arc exposures corrected by sky-emission lines, and by optimally subtracting the sky-emission lines. Optimal extraction of the spectral traces in each on-source exposure was then performed using the trace of the standard star Feige 67, which was also used for flux calibration and the removal of telluric absorption lines. The final optimally co-added spectrum of PSO J207+72 has a flux calibration uncertainty of 10% due to the differing airmasses of the standard-star and source observations. The galaxy was only significantly detected in the red arm of LRIS, and a truncated spectrum (displayed in Figure 3) was used for further analysis. In addition to the [OII] 3727 Angstrom emission-line doublet, some significantly detected absorption lines (CaII H and K lines at 3935 Angstrom and 3970 Angstrom, respectively; Hγ at 4342 Angstrom and H$\delta$ at 4103 Angstrom; the Fraunhofer G feature at 4306 Angstrom) are labelled in Figure 3. All lines were detected at a redshift of 0.660(2).

**Modelling of the host galaxy.** The Pan-STARRS photometry and KeckI/LRIS spectrum of PSO J207+72 was modelled using the Prospector code for stellar population inference. Prospector enables Markov Chain Monte Carlo (MCMC) sampling of the posterior distribution of parameters of the stellar populations and star-formation histories of galaxies, given a combination of photometric and spectroscopic data. Galaxy emission is modelled using a wrapper to the Flexible Stellar Population Synthesis code[42,43]. We fit a five-parameter `delay-tau' model for the stellar population of PSO J207+72, including the metallicity, the stellar population age and star-formation timescale, the mass in formed stars, and the *V*-band extinction of a dust screen. Prior to performing the fit, the observations were corrected for Galactic extinction using the 'extinction' software package[44], through a standard Milky Way extinction curve with a *V*-band extinction of 0.052 magnitudes. Data surrounding the detected [OII] 3727

Angstrom emission-line doublet were masked, and no modelling of nebular emission was conducted. Exploration of the posterior parameter distributions was conducted using the emcee MCMC software[45]. Standard Prospector priors were implemented. We derived a metallicity of 0.3(2) of the solar metallicity, a mass in formed stars of $10^{11.07(6)}\,M_{\odot}$, an age of 6.6(8) Gyr, a star-formation timescale of 1.0(2) Gyr, and a *V*-band extinction of 0.3(2) magnitudes.

PSO J207+72 lies within what appears to be a group of galaxies (Figure 2) with Pan-STARRS *r*-band magnitudes ranging between 19 and 23. No spectra are currently available for galaxies within this group, and the association in distance cannot therefore be confirmed. PSO J207+72 is undetected in observations from the Very Large Array Sky Survey[46]; the upper limit at 3 GHz on any source within the 99% confidence containment region of FRB 190523 is 0.36 mJy ($3\sigma$). Throughout this Letter, we use cosmological parameters from the 2015 Planck analysis[47].

**Data availability statement.** The datasets generated during and/or analysed during the current study are available from the corresponding author on reasonable request.

**Code availability statement.** Custom code is made available at https://github.com/VR-DSA.

**Extended Data Table 1 | ITRF coordinates of the ten DSA-10 antennas.**

| Antenna | X (m) | Y (m) | Z (m) |
|---|---|---|---|
| 1 | -2409464.509 | -4477971.270 | 3839125.031 |
| 2 | -2409466.445 | -4477974.866 | 3839119.657 |
| 3 | -2409470.315 | -4477982.059 | 3839108.908 |
| 4 | -2409547.552 | -4478125.603 | 3838894.418 |
| 5 | -2409468.38 | -4477978.463 | 3839114.282 |
| 6 | -2409429.957 | -4478294.47 | 3838772.061 |
| 7 | -2409682.474 | -4478158.598 | 3838772.061 |
| 8 | -2409746.758 | -4478124.008 | 3838772.061 |
| 9 | -2409770.667 | -4478111.143 | 3838772.061 |
| 10 | -2410525.007 | -4477850.573 | 3838597.062 |

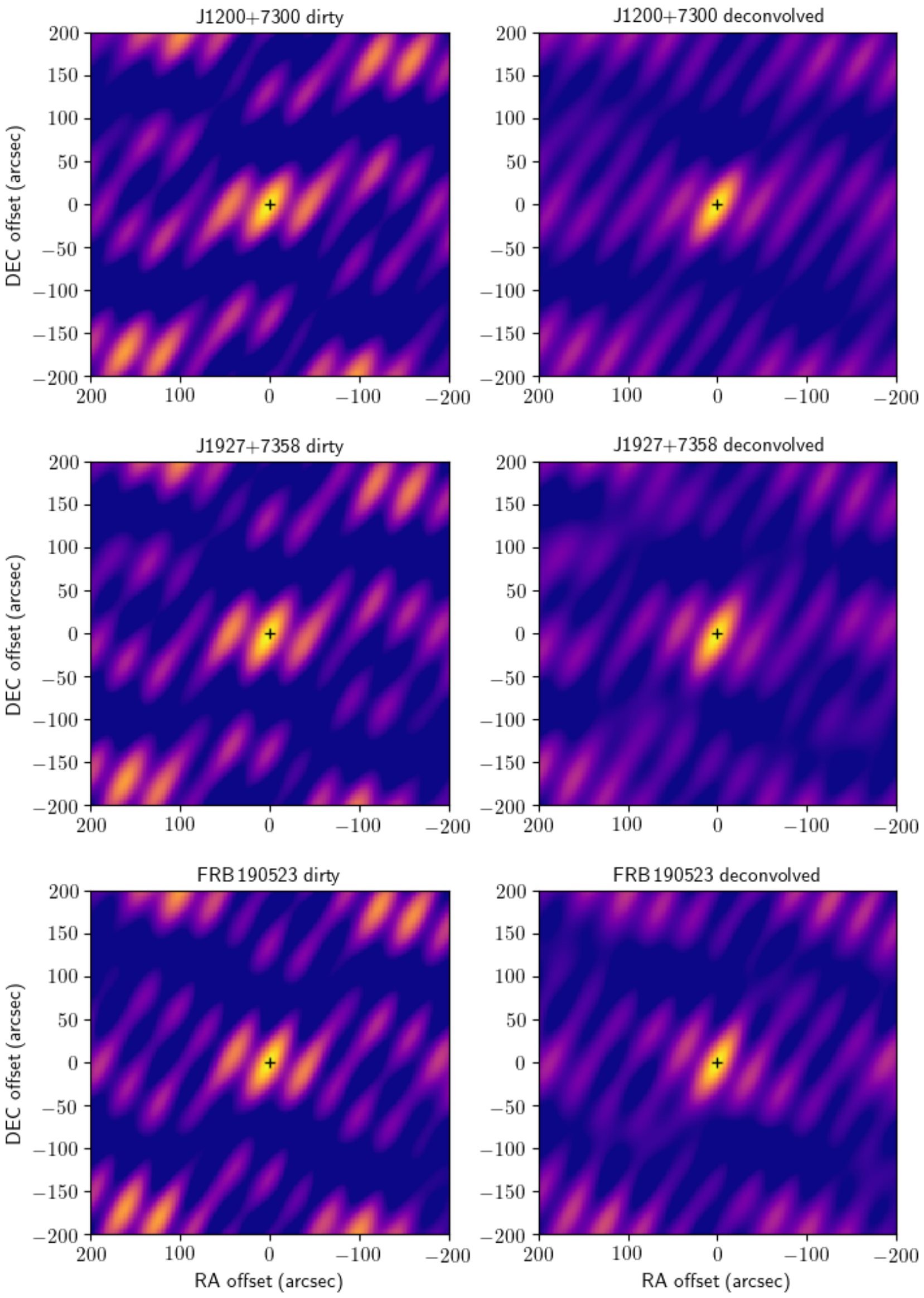

**Extended Data Figure 1 | DSA-10 images.** Dirty and deconvolved images are shown of two bright point-sources and FRB 190523. All data were obtained at the same hour angle relative to the meridian, within 12 hours of each other. The same calibration solution, partially derived using the J1200+7300 data, was applied to all data. The black crosses indicate the known source positions in the top and middle rows, and the best-fit position of FRB 190523 in the bottom row. The recovery of the correct position of J1927+7358 at the hour angle that FRB 190523 was detected at demonstrates the accuracy of the calibration solutions.

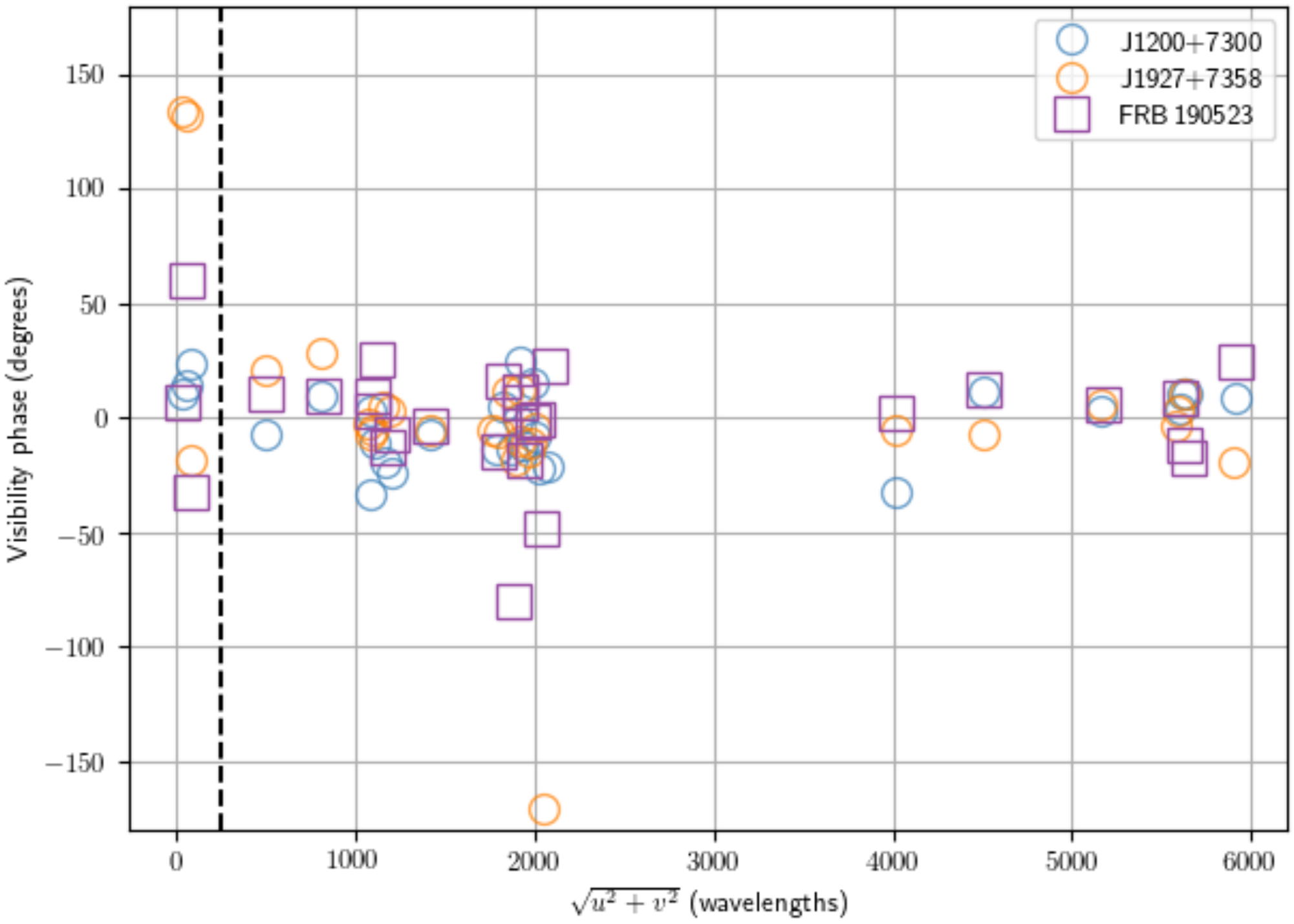

**Extended Data Figure 2 | Visibility phases measured for two bright point-sources and FRB 190523.** Only data on baselines with fully functioning antennas are shown. The visibility data were phase-rotated to the known (or best-fit for FRB 190523) source positions, and averaged across the frequency band. Data on the shortest baselines (to the left of the dashed vertical line) were corrupted by correlated noise, and were discarded from imaging analysis. All data were calibrated using the same calibration solution, which was partially based on the J1200+7300 data, and were obtained at the same hour angle relative to the meridian within a 12-hour timeframe.

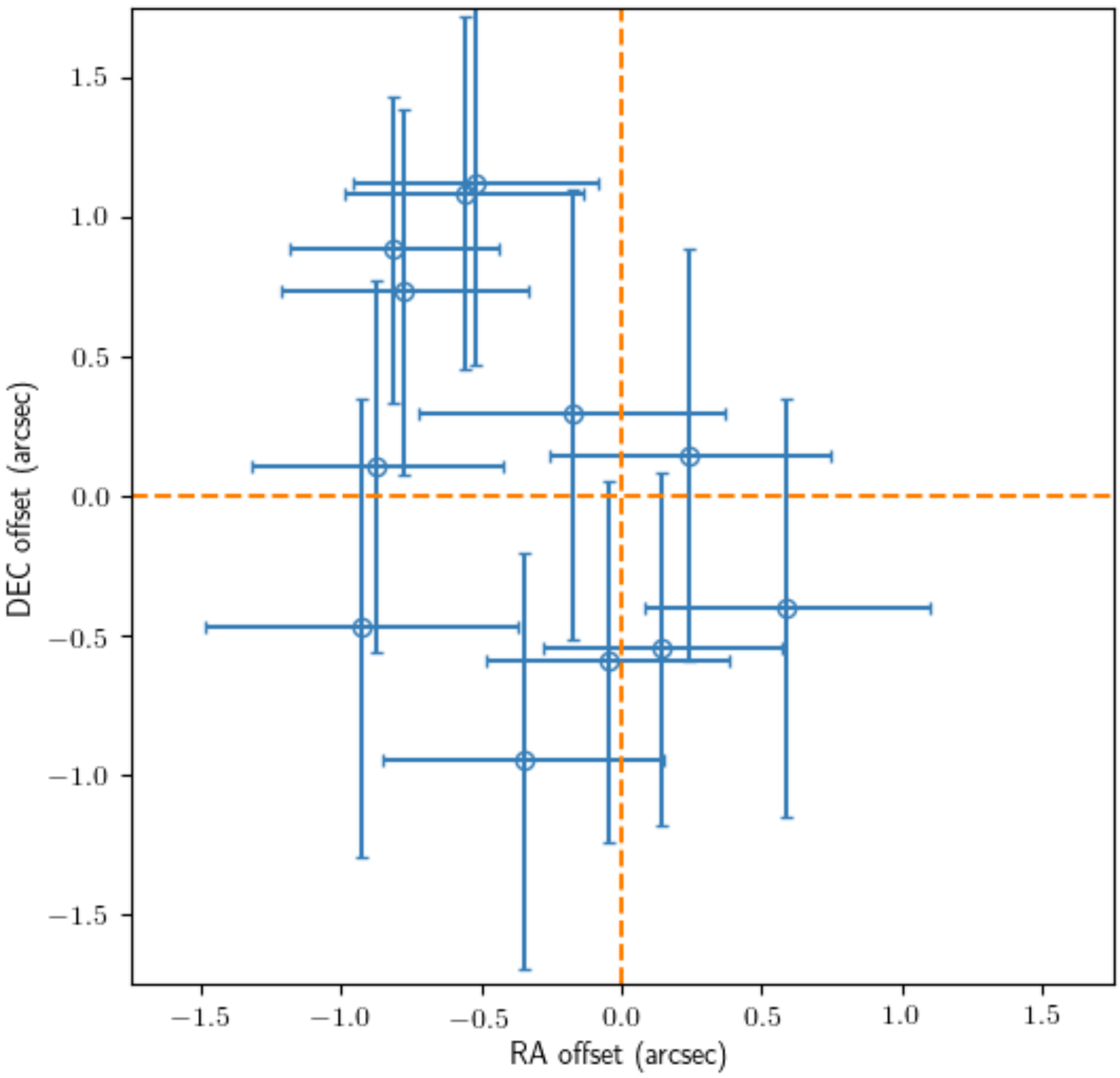

**Extended Data Figure 3 | Recovered positions of J1927+7358 on 12 separate days.** Each position was derived from 5 min of visibility data extracted when J1927+7358 was at the same hour angle as FRB 190523 was detected. On each day, the data were also calibrated exactly as the FRB 190523 data. The error bars indicate the 68% (1σ) confidence intervals for the estimated positions.